\documentclass[fleqn,usenatbib]{mnras}

\usepackage{newtxtext,newtxmath}

\usepackage[T1]{fontenc}
\usepackage{ae,aecompl}

\usepackage{xfrac}
\usepackage{bm}
\usepackage{hyperref}
\usepackage{booktabs}
\usepackage[inline]{enumitem}

\usepackage{graphicx}	
\usepackage{amsmath}	
\usepackage{amssymb}	

\newcommand{\Jcirc}{J_\text{circ}}

\newcommand{\jzeff}{j_{z,\text{eff}}}
\newcommand{\jzeffmean}{\bar{j}_{z,\text{eff}}}
\newcommand{\djzeff}{\Delta j_{z,\text{eff,max}}}
\newcommand{\esa}{\epsilon_\text{SA}}
\newcommand{\hierarchy}{r_{p,\text{out}} / a}

\title[Extreme close approaches in triple systems]{Extreme close approaches in hierarchical triple systems with comparable masses}

\author[N. Haim et al.]{
Niv Haim\thanks{E-mail: niv.haim@weizmann.ac.il},
Boaz Katz
\\
Weizmann Institute of Science, Rehovot, Israel\\
}

\date{Accepted XXX. Received YYY; in original form ZZZ}

\pubyear{2018}

\begin{document}
\label{firstpage}
\pagerange{\pageref{firstpage}--\pageref{lastpage}}
\maketitle

\begin{abstract}
We study close approaches in hierarchical triple systems with comparable masses using full N-body simulations, motivated by a recent model for type Ia supernovae involving direct collisions of white dwarfs (WDs). For stable hierarchical systems where the inner binary components have equal masses, we show that the ability of the inner binary to achieve very close approaches, where the separation between the components of the inner binary reaches values which are orders of magnitude smaller than the semi-major axis, can be analytically predicted from initial conditions. The rate of close approaches is found to be roughly linear with the mass of the tertiary. The rate increases in systems with unequal inner binaries by a marginal factor of $\lesssim 2$ for mass ratios ${0.5\leq m_1/m_2 \leq1}$ relevant for the inner white-dwarf binaries. For an average tertiary mass of $\sim 0.3 M_{\odot}$ which is representative of typical M-dwarfs, the chance for clean collisions is $\sim 1$\% setting challenging constraints on the collisional model for type Ia's. 
\end{abstract}

\begin{keywords}
gravitation -- celestial mechanics -- white dwarfs -- supernovae: general -- planets and satellites: dynamical evolution and stability
\end{keywords}



\section{Introduction}

The progenitor problem of type Ia supernovae (SNe Ia) - what triggers the thermonuclear explosion of some white-dwarfs (WDs)? remains open despite decades of observational and theoretical research \citep[see e.g.][for theoretical and observational reviews]{hillebrandt_Ia_review_2000,maoz_observational_2014}. It was recently suggested that a primary channel for SNe Ia may be the direct collisions of WDs \citep{kushnir_head-collisions_2013,katz_rate_2012,dong_type_2015}. A successful explosion in such collisions has been established and several simple and robust properties have been shown to agree with SNe Ia observations \citep{rosswog_collisions_2009,raskin_56ni_2010,hawley_zero_2012,kushnir_head-collisions_2013,dong_type_2015}. 
 
The main challenge of this model is whether the collision rate is sufficient to account for the type Ia rate, which requires about $1\%$ of WDs to explode \citep[e.g.][]{maoz_star_2017}. The collision rate of free floating WDs in the field is off by many orders of magnitudes. Collisions in the dense cores of globular clusters are insufficient given the small $10^{-4}$ fraction of stars that reside in these systems \citep[e.g.][]{rosswog_collisions_2009}. Recent advances in the study of the long term evolution of few-body systems \cite{ford_secular_3body_2000,naoz_hot_2011,katz_long-term_2011,bode_production_2014,antonini_secular_2012,katz_rate_2012,pejcha_quadruple_2013} have lead to the realization that extremely close approaches often occur in such systems. Close approaches between WDs in multiple stellar systems may result in mergers due to the significant gravitational wave emission \citep{thompson_accelerating_2011} or direct collisions \citep{thompson_accelerating_2011,katz_rate_2012}. If a sufficient amount of WDs are in relevant multiple systems, the rate of direct collision may be as high as the type Ia rate \cite{katz_rate_2012}. 

Is the collision rate sufficient to account for the the type Ia rate? An estimate of the collision rate requires the knowledge of the multiplicity properties of WDs and an understanding of the dynamics of multiple systems. Both of these aspects are not sufficiently known to provide a reliable answer at the moment \citep[note claims to the contrary and that the rate is much too low][]{hamers_population_2013,soker_11fe_2013,toonen_rate_2017}. WDs in multiple systems are hard to observe since their main-sequence companions greatly outshine them \citep[e.g.][]{Holberg_sirius_like_2013,katz_missingWDs_2014}. In particular, the closest known WDs - Sirius B and Procyon B, would probably not have been discovered had they been a few times farther away, and it is likely that the majority of such systems have not been discovered yet even within the local solar neighborhood \citep{ferrario_missingWDs_2012,katz_missingWDs_2014}. Note that the question of how many WDs have an additional WD companion is much more well constrained and it is estimated that about $10\%$ of WDs have a lighter WD companion on a relevant wide-orbit ($0.5< period <5000$yr)\citep{klein_way_2017,maoz_doubleWDs_2017}. A significant improvement in the census of WDs in multiples is expected in the near future by the now operating European astrometric mission Gaia. 

On the theoretical side, the WD collision model faces the following challenge - if a given system is sufficiently active dynamically as to lead to a collision of the WDs, why didn't it lead to the collision of the main sequence progenitors of the WDs which are much bigger \citep{katz_rate_2012}? Possible ways around this can be related to changes in the system's configuration occurring on time scales which are longer than the stellar evolution time due to passing stars \citep[e.g.][]{antognini_clusters_2016} or multiplicity which is higher than 3 \citep[e.g.][]{pejcha_quadruple_2013,fang_dynamics_2017} or to small changes in the configuration in the last stages of mass loss leading to the formation of the WDs.
Indeed, population synthesis calculations of isolated triple systems, assuming isotropic, adiabatic mass loss have obtained very low collision rates \citep{hamers_population_2013,toonen_rate_2017}. While this is a serious challenge, our poor knowledge of the primordial systems and mass loss processes do not allow clear conclusions. 

In this paper we focus on the simplest dynamical aspect of the collision model which has only been partly explored - the conditions for close approaches in triple systems after the inner binary stars became WDs. We use direct numerical n-body simulations to map the conditions for close approaches for the relevant parameter space of configurations and mass values of the 2 WDs and the third star. This work extends the results presented in \cite{katz_rate_2012} in three important ways: First, the dependence on the mass is studied down to low tertiary masses of $0.1 \rm M_{\odot}$ which are relevant to the abundant M-dwarfs. Second, we derive analytic criteria that allow the collision probability to be estimated with good confidence for the case of nearly equal mass WDs. Third, we study the dependence on the mass ratio of the inner-binary. For the purpose of this work, a new code was written in \textit{Python} by N. Haim which is now publicly available.  

The structure of this paper is as follows. In section~\ref{sec:direct_integration} we describe the numerical integrations that are performed and show the main numerical results. In section~\ref{sec:numerical_example} we show the evolution in a few numerical examples  and emphasize the main characteristics of triple systems leading to extreme close approaches. In section~\ref{sec:ca_criterion} we derive an analytic criterion for extreme close approaches in systems with equal-mass inner binaries, and in section~\ref{sec:sneia_implications} we summarize the results and discuss the implications for the collision model of type Ia's.

\section{Direct numerical integrations show that the close-approaches rate is roughly linear with \texorpdfstring{$\MakeLowercase{m}_3$}{m 3}}
\label{sec:direct_integration}

We consider \textit{hierarchical} triple systems of three gravitationally interacting bodies of masses $m_1,m_2,m_3$, where the distance from $m_3$ to either $m_1$ or $m_2$ is much larger than the distance between $m_1$ and $m_2$. On short time scales the hierarchical triple system behaves as two 2-body (Keplerian) systems: the smaller \textit{inner orbit} of $m_1$ and $m_2$ and the larger \textit{outer orbit} of $m_3$ and the center of mass of $m_1$ and $m_2$. There is very little exchange of energy between the two orbits and therefore both inner and outer semi-major axes ($a$ and $a_{out}$) are almost constant. Accordingly the hierarchical arrangement of the system stays stable on very long timescales. In contrast, there is an exchange of angular momentum. The orbital parameters other than the semi-major axis, and in particular the eccentricity, change slowly on time-scales much larger than both orbital periods $P$ and $P_{out}$.

It is useful to quantify the \textit{hierarchy} of the system as the initial ratio between the pericenter of the outer orbit $r_{p,out}$ and $a$, namely the minimal separation between $m_3$ and the center of mass of the inner binary, expressed in units of $a$. Systems with small hierarchy experience substantially more exchange of energy and they may be quickly disrupted, with one of the masses ejected to infinity. On the other hand, systems with higher hierarchy stay stable for longer periods. For example, systems of three comparable masses with hierarchies larger than 3-4 are stable for millions of orbits \citep[e.g.][]{he_collisions_2018}.

In this paper we focus on hierarchical triples with moderate hierarchy $1.5 < r_{p,out} / a < 10$. The inner binary includes two White Dwarfs (WD) of mass $0.6 < m_1, m_2 < 1.2$ $M_{\sun}$ with semi-major axis $1 < a < 1000 AU$. The perturber is a stellar object of mass $0.1 < m_3 < 1.2$ $M_{\sun}$. Most of our runs include an equal mass binary $m_1=m_2=0.6M_{\sun}$ (where there is a peak in the WD mass distribution, e.g. \cite{holberg_25_2016}) but we also perform simulations with inner binaries having unequal masses.

We study the conditions for extreme close approaches in the inner binary, where the inner binary separation ($r=r_1-r_2$) becomes smaller by orders of magnitude compared to the semi-major axis $a$. A system is said to have experienced a close approach if $r / a$ becomes smaller than a pre-defined threshold. For example, in order that two white dwarfs in an orbit with $a=10$AU experience a collision, the separation has to be smaller than twice the white-dwarf radius or $r/a < 2R_{WD}/10 \text{AU} \sim 10^{-5}$.

\subsection{Numerical simulations}
\label{sec:numerics}

Our goal is to estimate the probability of achieving close approaches. This is a statistical property which can be estimated based on large ensembles of integrations with varying initial conditions. 

We run $\sim$120,000 full N-body simulations of three bodies using a second-order symplectic integrator with an adaptive time-step \cite{preto_class_1999,mikkola_algorithmic_1999} (PTMT). Details of the integrator are provided in appendix~\ref{app:integration} and in \cite{katz_rate_2012}.  

\textit{Ensembles} Multiple runs are prefomred for ensembles of $\sim 500-2000$ randomly chosen initial conditions (see Table~\ref{tab:sim_ensembles} for exact details). All the simulations in a given ensemble have the same mass values and initial $\hierarchy$.  The initial orbital parameters of each system in an ensemble are chosen randomly from the following distribution:

\textit{Initial conditions} Both the inner and the outer eccentricities are chosen uniformly at random in the range $0 < e, e_{out} < 0.9$. The z-axis is chosen along the initial direction of the outer angular momentum and the x-axis along the outer orbit's eccentricity vector (Runge-Lenz vector). The orbital orientation of the inner orbit is chosen randomly from an isotropic distribution, with uniform distributions of $0 < \Omega,\omega < 2\pi$ and $-1 < \cos{i} < 1$, where $\Omega$ is the longitude of ascending node, $\omega$ is the argument of periapsis and $i$ is the (mutual) inclination. The mean anomalies are chosen uniformly at random in $[0,2\pi]$. 

\textit{Stopping conditions} The evolution of the system is stopped if either
\begin{enumerate}[label=(\alph*)]
  \item $t / P_0 > 2 \times 10^6$, where $P_0$ is the (initial) inner period.
  \item The system is disrupted. The disruption condition is that the outer orbit has a positive energy and that its separation is more than 50 times larger than the separation of the inner orbit.
  \item In section~\ref{sec:sneia_implications} we add a condition and stop if 5 Gyr have passed. Note that each run can be scaled in distance and time, so that this condition is enforced in post-analysis.
\end{enumerate}

For each ensemble we estimate the close approach probability as the fraction of systems that experienced a close approach out of the total number of systems in the ensemble. In section \ref{sec:sneia_implications} we discuss clean collisions, in which only systems that did not have close approaches prior to the collision are counted \cite[e.g.][]{katz_rate_2012,he_collisions_2018}.  
Convergence is demonstrated in Appendix~\ref{app:integration}.

We provide the simulation code, written in $\textbf{Python}$. The performance is enhanced using the numba package for generating just-in-time machine instructions. The code can be found here: \url{https://github.com/nivha/three_body_integration}

\subsection{Direct estimates of the close approach probability}

\begin{figure}
	\includegraphics[width=\columnwidth]{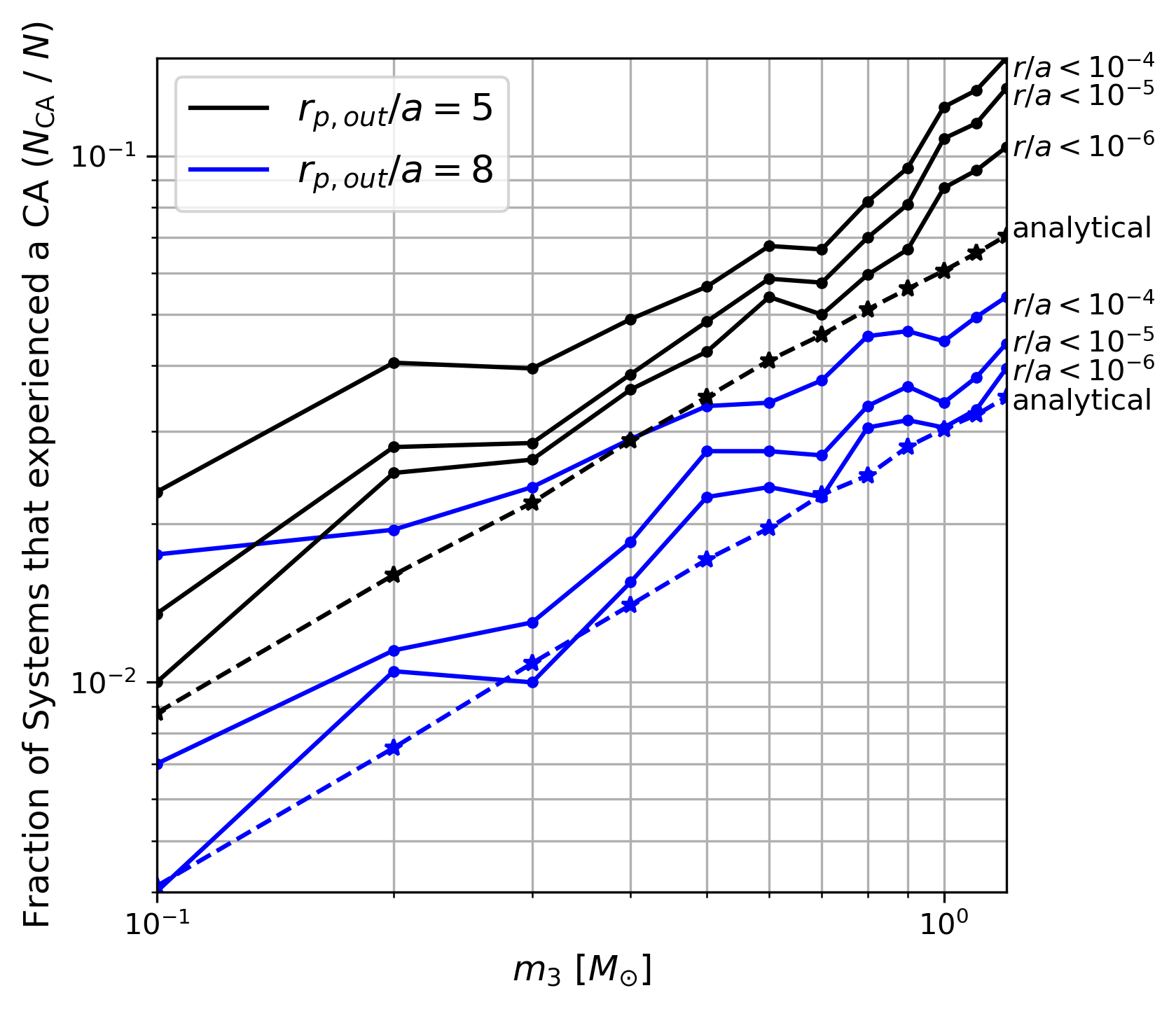}
    \caption{Extreme close approaches (CA) for triple systems with equal mass inner binaries. Each point represents the result of an N-body integration of an ensemble of $\sim 500$ three body systems with random initial conditions as described in section \S\ref{sec:numerics} and table~\ref{tab:sim_ensembles}, integrated to $2\cdot10^6$ inner periods. The y-axis shows the fraction of systems that experienced a close approach out of the total number of systems for that ensemble. The x-axis shows the mass of the tertiary ($m_3$). The systems have inner binary of masses $m_1=m_2=0.6M_{\sun}$ with moderate hierarchy $\hierarchy=5,8$ (black, blue). Results are shown for close approach thresholds of $r/a<10^{-4},10^{-5},10^{-6}$ as indicated in the figure. The dashed lines are the expected asymptotic fractions for $r/a\rightarrow 0$, from an analytic approximation derived in section~\S\ref{sec:ca_criterion} and applied to a large sample of initial conditions with the same distribution.}
    \label{fig:rates_per_mp}
\end{figure}

Figure~\ref{fig:rates_per_mp} shows the close approaches probability as a function of $m_3$. The systems shown have masses $m_1=m_2=0.6M_{\sun}$ and $0.1 < m_3 < 1.2 M_{\sun}$ and initial hierarchies of $\hierarchy=5$ (black) and $\hierarchy=8$ (blue). We show the results for close approach thresholds of $r/a < 10^{-4},10^{-5},10^{-6}$. The dashed line is the analytic expected asymptotic rate of $r/a\rightarrow0$ which is derived in section~\ref{sec:jzeff_fluct}, and applied to initial conditions of $500,000$ systems sampled from the same distributions as the numerical simulations.

\begin{figure}
	\includegraphics[width=\columnwidth]{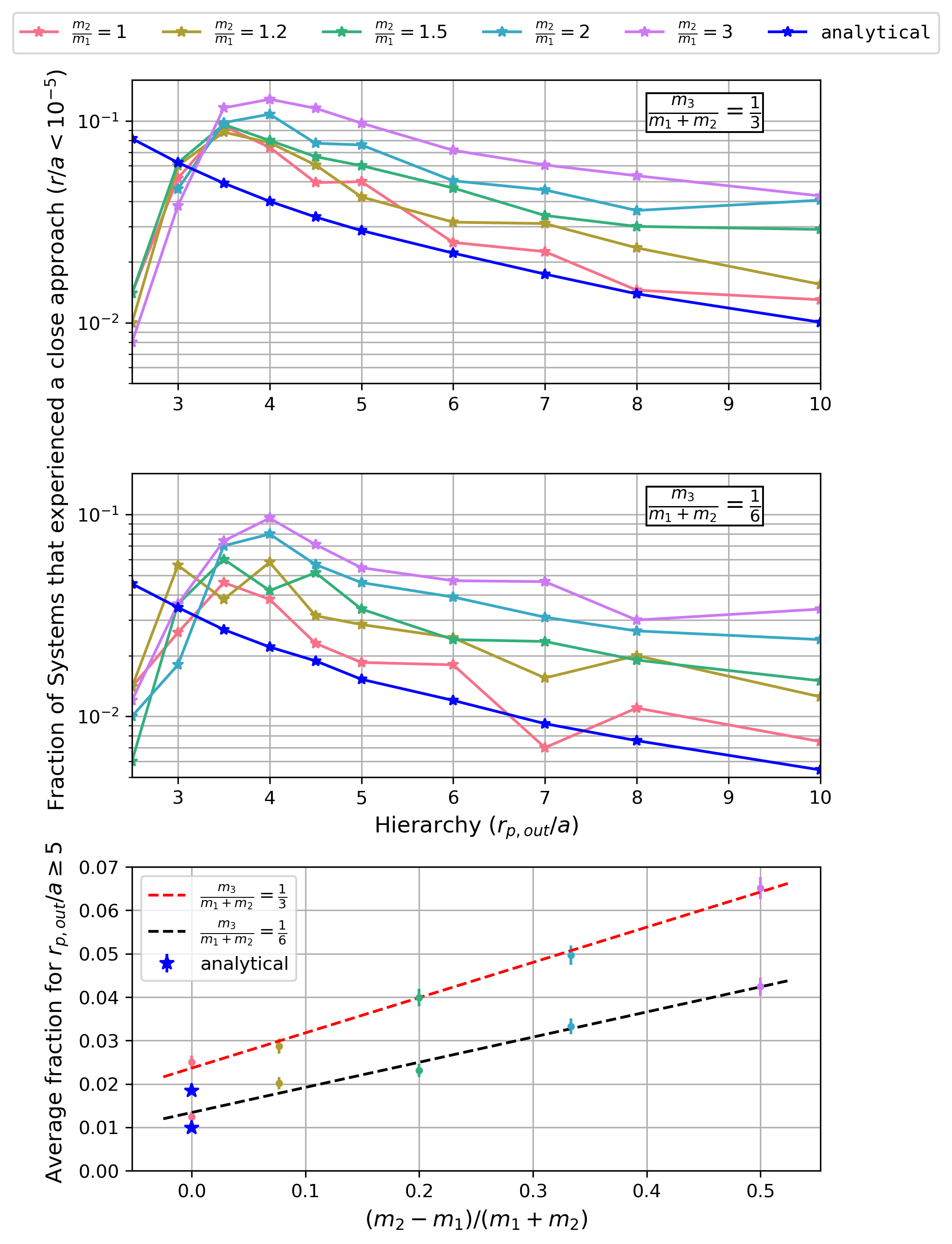}
    \caption{Extreme close approaches for systems with unequal inner binaries. \textit{Top and middle panels}: Each point represents the result of an ensemble of $\sim 500$ three body system integrations as in figure~\ref{fig:rates_per_mp}. The x-axis shows the hierarchy $\hierarchy$ of the system. The close approach threshold is set to $r/a=10^{-5}$. Results are shown for systems with inner binary mass ratio $m_1/m_2=1,\frac{7}{6}, \frac{3}{2}, 2, 3$ and third body mass ratios, quantified as $m_3/(m_1+m_2) = \frac{1}{3},\frac{1}{6}$ (upper and middle panels resp.). \textit{Lower panel}: Each point is the average probability (and statistical error) obtained in the simulations with  $5\leq\hierarchy\leq10$ shown as a function of the octupole mass coefficient $(m_2-m_1)/(m_2+m_1)$.  The dashed lines are linear fits to the points. 
The analytical average rate for equal mass binaries is shown in blue stars in all panels (connected by lines in the top and middle panels).}
    \label{fig:col_rates_octupole}
\end{figure}

As can be seen in Figure~\ref{fig:rates_per_mp}, the close approaches probability increases with smaller initial hierarchy. The rate is insensitive to the exact value of the close approach threshold for the relevant extreme values $r/a\lesssim 10^{-5}$ and is close to the analytic asymptotic value. The rate decreases with the mass of the perturber $m_3$ roughly linearly. 

Figure~\ref{fig:col_rates_octupole} shows the close approach probability as a function of the hierarchy $\hierarchy$ for a threshold of $r/a<10^{-5}$. Results are shown for different mass choices including three inner-binary mass ratios $m_2/m_1=1,\frac{7}{6}, \frac{3}{2}, 2, 3$, and two third-body mass ratios, quantified as $m_3/(m_1+m_2) = \frac{1}{3},\frac{1}{6}$ (upper and middle panels resp.). 
 
As can be seen in Figure~\ref{fig:col_rates_octupole}, for the mass ranges that we consider, the dependence of the close approaches probability on the inner-binary mass ratio is not very large; For a rare case of WD-WD binary of mass ratio $m_2/m_1=2$ the probability is about twice as large as the rate for equal mass binary with similar perturber. For the more probable masses of WD-WD binaries, this difference will be even smaller.

The close approach probability is suppressed at small hierarchies due to the quick disruption of the systems and at large hierarchies due to the smaller range of initial conditions leading to high eccentricities.
We next focus on the dependence of the close approach probability at higher hierarchies. For a study of the effects of disruptions on the close approach rates see \citep{he_collisions_2018}.

\section{Numerical examples of a triple systems}
\label{sec:numerical_example}

\begin{figure}
	\begin{tabular}{c}
	\includegraphics[width=\columnwidth]{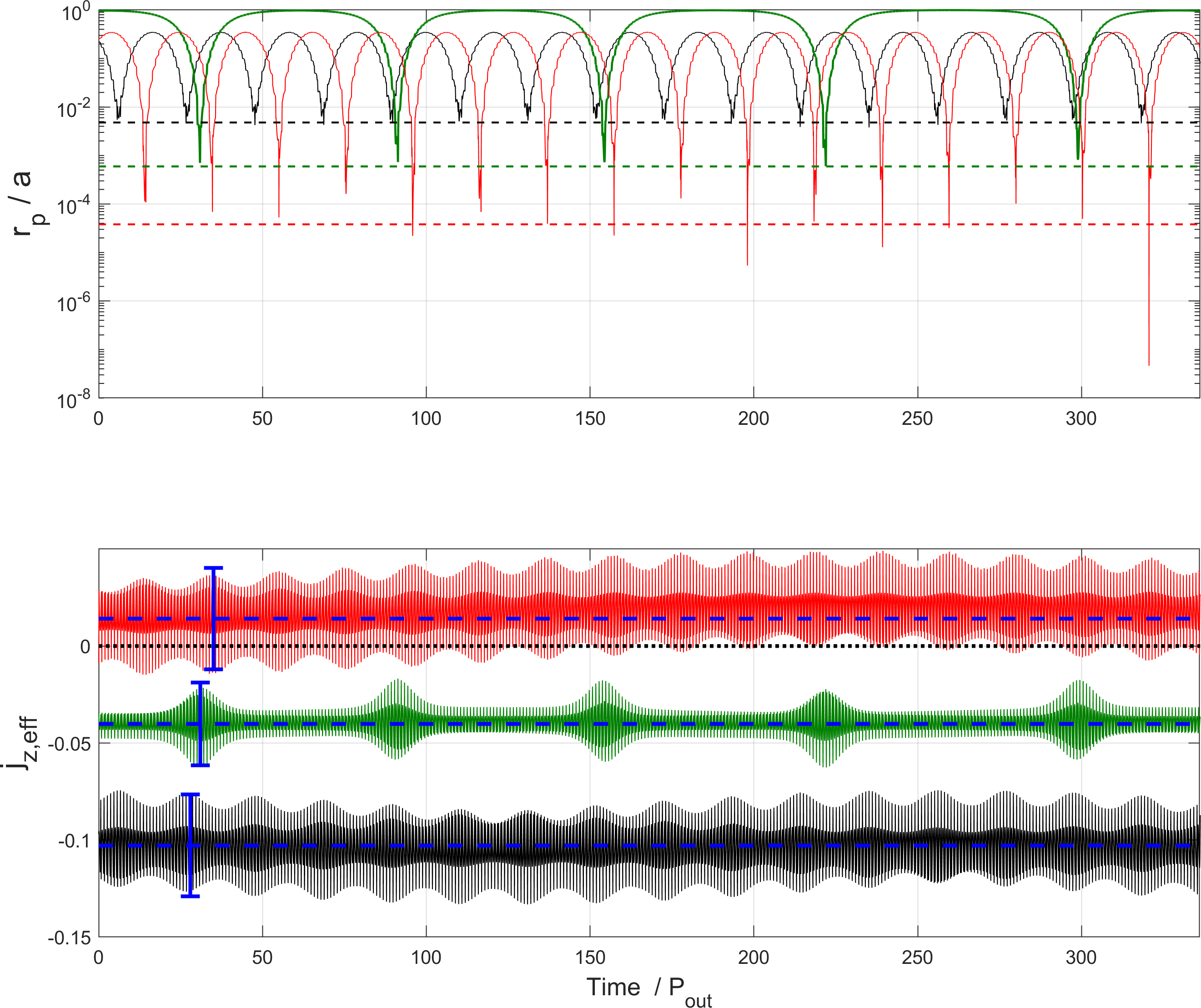} \\
    \begin{tabular}{llllllll}
    \toprule
     Color &   $e$ & $e_{out}$ &   $i$ & $\Omega$ & $\omega$ & $M_{in}$ & $M_{out}$ \\
    \midrule
       Red &  0.78 &      0.38 &  1.79 &     1.46 &      3.5 &      1.7 &      2.21 \\
     Green &  0.03 &      0.32 &  1.71 &     6.11 &     2.86 &     2.55 &      0.82 \\
     Black &  0.76 &      0.44 &  1.65 &      3.1 &     2.82 &     2.63 &      6.28 \\
    \bottomrule
    \end{tabular}
    \end{tabular}
    \caption{Examples of triple systems with equal mass inner binaries  (three integrations shown in red, green, black). All three systems have masses $m_1=m_2=m_3=0.6M_{\sun}$ and initial hierarchy $\hierarchy=7$. \textit{Upper panel}: the evolution of the inner orbit pericenter. Dashed lines are the minimal values for the pericenter as predicted by the double-averaged, quadrupole approximations (Kozai-Lidov oscillations, eq.~\ref{eq:J_bound}). The red system breaks the rule and achieves minimal separations which are smaller than the predicted value (dashed) by orders of magnitude. \textit{Lower panel}: the evolution of $\jzeff$ which fluctuates around a constant mean value (dashed blue). The minimal value of $\lvert \jzeff \rvert$ sets a lower bound for the inner angular momentum (eq.~\ref{eq:j_bound}) and consequently to the minimal separation (eq.~\ref{eq:rmin_lk}). If $\jzeff$ crosses zero (as in the red system) then $j$ is no longer lower-bounded and the inner separation can become arbitrarily small, as seen in the upper panel. Both mean values (dashed blue) and bounds (blue bars) of $\jzeff$ are computed from the initial conditions of the system (see eq.~\ref{eq:jzeff_djzeff},~\ref{eq:djzeff_max}), this means that the ability of $\jzeff$ to cross zero, and therefore achieving very close approaches, can be predicted from initial conditions. The table shows the initial conditions for the systems (inner and outer eccentricities, mutual inclination, longitude of ascending node, argument of periapsis, inner and outer mean anomalies. Angles are in radians).}
    \label{fig:evolution}
\end{figure}

In Figure~\ref{fig:evolution} are shown examples for the evolution of the orbital parameters in hierarchical triples. All three systems (shown in black, green and red) have masses $m_1=m_2=m_3=0.6M_{\sun}$ and initial hierarchy $\hierarchy=7$.

The evolution of the pericenter of the inner binary $r_p$ relative to the (almost constant) inner semi-major axis $a$ is shown in the upper panel. As can be seen it is possible for $r_p/a$ to reach very small values of $<10^{-2}$ and that the behavior is approximately periodic. The long term periodic evolution of hierarchical triple systems is well explained by the Lidov-Kozai (LK) mechanism (\citet{lidov_evolution_1962,kozai_secular_1962}) and can be approximated using the following basic assumptions:
\begin{enumerate}[label=(\alph*)]
  \item There is no exchange of energy between the inner and outer orbit ($a,a_{out}=const.$).
  \item The Hamiltonian is separated to the two Keplerian Hamiltonians of the inner and outer orbits and a small perturbing potential which is expanded up to quadratic terms as a function of the small ratio between the inner and outer separations.
  \item The small changes in the Keplerian orbital parameters (e.g. $\Delta j$, and $\Delta e$) due to the perturbation are averaged twice: once over one outer period and once over one inner period (this is called \textit{double averaging}).
\end{enumerate}

Under these assumptions, the double averaged quadratic potential turns out to be axisymmetric with respect to the direction of the outer angular momentum $\hat{\textbf{J}}_{out}$. Therefore no torques exist along this axis, and the magnitude of the outer angular momentum is conserved:

\begin{equation}
	\lvert J_{out} \rvert = const.
\end{equation}

One consequence is that the number of constants turns out to be one less than the number of degrees of freedom and the resulting long-term evolution is periodic. Another consequence is that $J$ has a lower bound set by the constants of motion. In fact, given that the total angular momentum $\lvert J_{tot} \rvert$ is conserved it follows that

\begin{equation}
	\lvert J \rvert > \Big\lvert \lvert J_{tot} \rvert - \lvert J_{out} \rvert \Big\rvert.
    \label{eq:J_bound}
\end{equation}

Equation \eqref{eq:J_bound} can be used to put a lower bound on the separation of the inner binary, which is shown in the upper panel of figure~\ref{fig:evolution} (as dashed lines). As can be seen, this lower bound works well for the systems shown in black and green but it fails for the integration shown in red. The reason for the failure is that in reality $J_{out}$ is fluctuating on short time-scales and these fluctuations are not taken into account by the double-averaging approximation \citep{bode_production_2014,antonini_secular_2012,katz_rate_2012,luo_double-averaging_2016}. 

In order to derive a condition for close approaches, the fluctuations in $J_{out}$ need to be quantified. Given that even small relative fluctuations in $J_{out}$ may be important it is more convenient to quantify the corresponding fluctuations in the inner-binary's angular momentum $\textbf{J}$. In particular in the test-particle limit, where one of the inner masses is negligible, the angular momentum vector of the outer orbit is fixed. In this case, the axisymmetry of the averaged perturbing potential implies that the component of the inner angular momentum along the axis of symmetry is constant. Here it is customary to choose the $z$ direction along the direction of the constant outer angular momentum vector. It is therefore useful (in the test-particle limit) to study the short-term fluctuations in $J_z$, which is fixed under the double-averaging approximation.

In the non test-particle case $\hat{\textbf{J}}_{out}$ is not conserved and it is useful to study the fluctuations in the following (double-averaging) constant which reduces to $J_z$ in the test particle limit (e.g. \citet{katz_rate_2012}): 
\begin{equation}
    \jzeff = \frac{\lvert J_\text{tot}\rvert^2 - \lvert J_\text{out}\rvert^2}{2\lvert J_\text{out}\rvert J_\text{circ}} = \textbf{j} \cdot \hat{\textbf{J}}_\text{out} + j^2 \frac{J_\text{circ}}{2\lvert J_\text{out}\rvert},
    \label{eq:jzeff}
\end{equation}
where
\begin{equation}
	\Jcirc = \mu \sqrt{G\left(m_1 + m_2 \right)a}
\end{equation}
is the value of the angular momentum of the inner orbit if it were circular (the maximal possible value),
\begin{equation}
	\textbf{j} = \frac{\textbf{J}}{\Jcirc}
    \label{eq:j_normalized}
\end{equation}
is the normalized inner angular momentum and $j=|\textbf{j}|$. Note that $j$ is directly related to the eccentricity through ${j^2 + e^2 = 1}$, and that at high eccentricities the pericenter can be expressed as:
\begin{equation}
	r_\textrm{peri} = \frac{1}{2} j^2.
    \label{eq:r_p_j2}
\end{equation}

In orbit with close approaches $j\ll1$ and ${\jzeff\approx \textbf{j}\cdot \hat{\textbf{J}}_{out}<j}$ (eq.~\ref{eq:jzeff}) implying a lower limit for $j$:
\begin{equation}
	j \geq \lvert \jzeff \rvert.
    \label{eq:j_bound}
\end{equation}
The lower limit Eq. \eqref{eq:j_bound} is equivalent to the lower limit \eqref{eq:J_bound} for close approaches where $|J|\ll|J_{tot}|$. To see this note that in this case, $|J_{tot}|\approx |J_{out}|$ so that ${|J_{tot}|^2-|J_{out}|^2\approx (|J_{tot}|-|J_{out}|)(2|J_{out}|)}$. Eq. \eqref{eq:j_bound} then follows from Eqs. \eqref{eq:jzeff} and \eqref{eq:J_bound}. By plugging $\jzeff$ into Eq.~\eqref{eq:r_p_j2} we obtain a lower limit for the separation of the inner binary:
\begin{equation}
	r_{min} = \frac{1}{2} \jzeff^2.
    \label{eq:rmin_lk}
\end{equation}
Note that the exact value of the minimal separation can be calculated using all the initial parameters and the conservation of the averaged perturbing hamiltonian \citep{lidov_evolution_1962,lidov_non-restricted_1976} and will be equal or higher than equation \eqref{eq:rmin_lk}.    

In the middle panel of Figure~\ref{fig:evolution} we show the evolution of $\jzeff$ for all three systems. First, note that $\jzeff$ is not strictly constant but rather fluctuating around a constant mean $\jzeffmean$ (shown in dashed blue). Second, note that for the system shown in red, the fluctuations in $\jzeff$ enable it to cross zero. In this case, $j$ is not bounded from below and the inner binary achieves very small separations, as is clearly seen in the upper panel. This is the major difference between the system shown in red and the other systems.

\begin{figure}
	\includegraphics[width=\columnwidth]{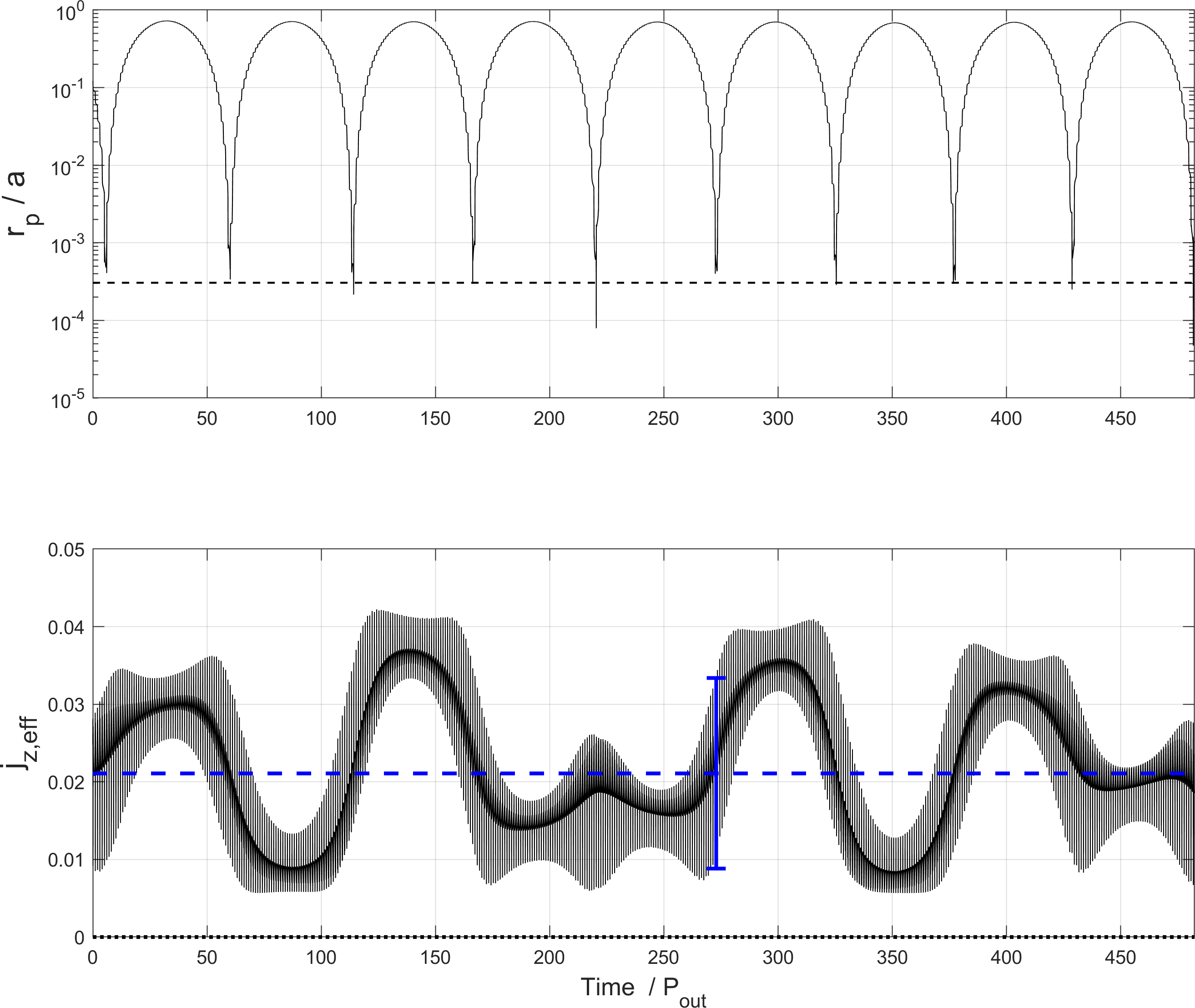}
    \caption{An example of the evolution of a triple system with an unequal mass binary. The system shown here is of masses $m_1=0.6,m_2=1.2,m_3=0.6M_{\sun}$ and initial hierarchy $\hierarchy=7$. The upper panel shows the evolution of the pericenter. The lower panel shows the evolution of $\jzeff$ (similar to Fig.~\ref{fig:evolution}). The octupole term of the perturbing potential (which vanishes for  equal mass inner binaries) causes the observed long term variations in $\jzeff$. The initial conditions for this system are: $e=0.87,e_{out}=0.81,i=1.62,\Omega=1.13,\omega=5,M_{in}=3.1,M_{out}=3.47$.}
    \label{fig:evolution_octupole}
\end{figure}

In Figure~\ref{fig:evolution_octupole} we show an example of the evolution of the pericenter and $\jzeff$ for the case of unequal binary masses ($m_1=0.6,m_2=1.2,m_3=0.6 M_{\sun}$). As can be seen $\jzeffmean$ is no longer constant. In this case, the quadratic expansion of the potential (assumption (b) of the LK mechanism) is not sufficient and the observed oscillations are caused by the next term, the octupole \citep{ford_secular_3body_2000,naoz_hot_2011,katz_long-term_2011,naoz_kozai_review_2016}. Due to these oscillations $\jzeff$ can cross zero and lead to extreme close approaches \citep[for the test particle case, this is equivalent to a 'flip' from pro-grade to retrograde orientations of the inner and outer orbits][]{naoz_hot_2011,katz_long-term_2011,luo_double-averaging_2016}. The octupole depends on the masses through the combination $\left(m_2-m_1\right) / \left(m_2+m_1\right)$ \citep[e.g.][]{naoz_kozai_review_2016}. As can be seen in the bottom panel of Fig~\ref{fig:col_rates_octupole} the close approaches probability increases linearly with the octupole term. Note that the short-term (outer period timescale) fluctuations have a long-term effect on the evolution \citep[][e.g. Fig.2]{luo_double-averaging_2016}. This is not taken into account in existing secular codes \citep[e.g.][]{naoz_hot_2011,hamers_population_2013,toonen_rate_2017,fang_dynamics_2017}, thus, calculating the evolution of $\jzeff$ requires more work.

\begin{figure}
	\includegraphics[width=\columnwidth]{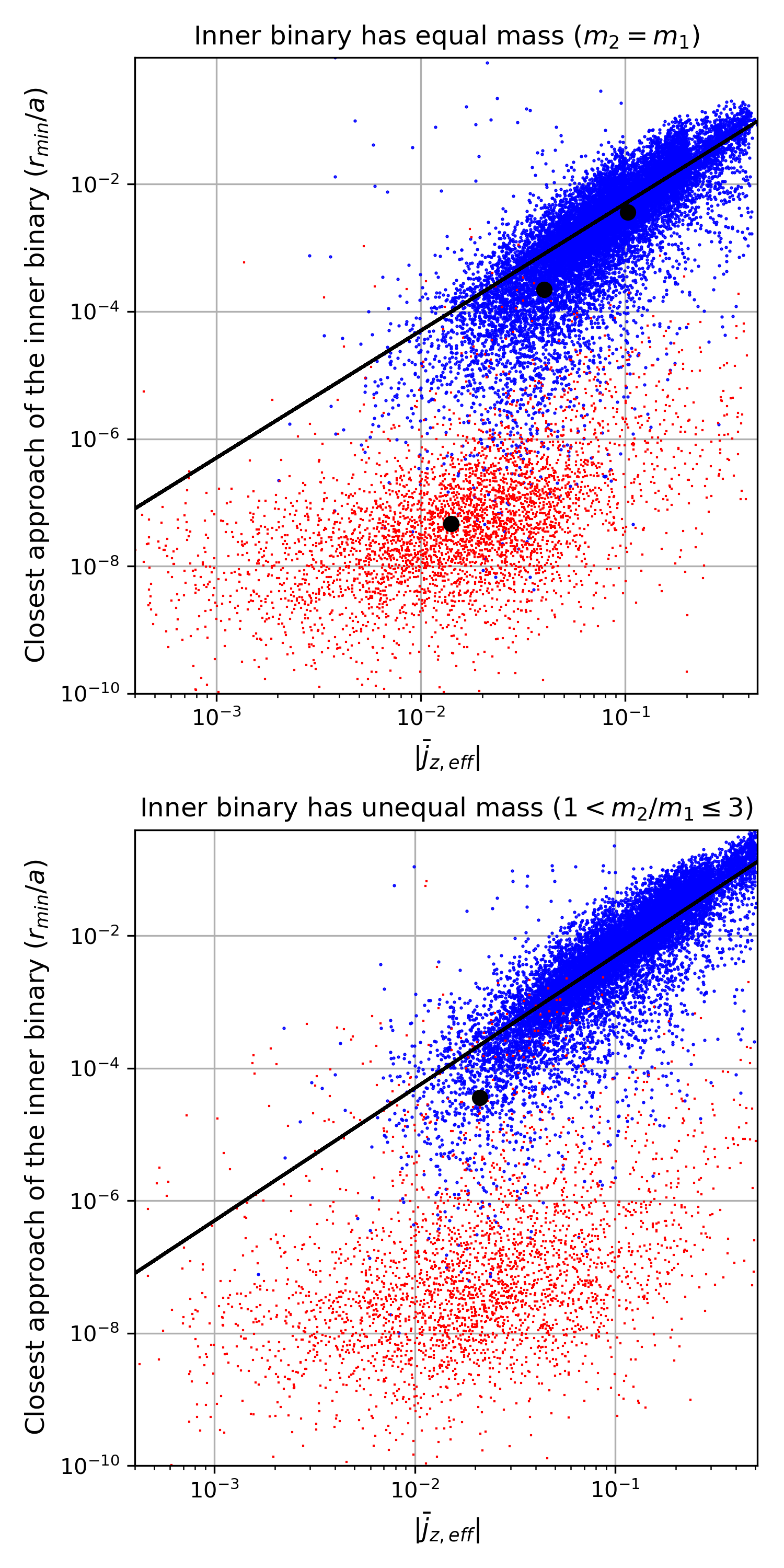}
    \caption{Close approaches in systems where $\jzeff$ crosses zero (red dots) verses those that do not (blue dots). Each dot is the result of a full n-body simulation from the ensembles presented in Figs. \ref{fig:rates_per_mp},\ref{fig:col_rates_octupole}, restricted to hierarchies  $5<\hierarchy<10$. 
    The x-axis is the initial value of $\jzeffmean$, and the y-axis is the minimal separation achieved by the inner binary throughout the evolution. The black line is the predicted minimal separation based on the Lidov-Kozai approximations (eq.~\ref{eq:rmin_lk}). \textit{Upper panel:} systems with equal mass inner binary with masses as in Fig~\ref{fig:rates_per_mp}, the three black dots are the three systems described in Fig~\ref{fig:evolution}. \textit{Lower panel:} systems with unequal mass inner binary (same masses as in fig.~\ref{fig:col_rates_octupole}) The black dot is the system described in Fig.~\ref{fig:evolution_octupole}. 
    \label{fig:rperi_vs_jzeff}}
\end{figure}

In Figure~\ref{fig:rperi_vs_jzeff} the initial values of $|\jzeffmean|$ and the minimal separations obtained are shown for all runs performed in this work. Each point in the figure represents the result of a triple system integrated to $2\cdot10^6$ inner periods. The y-axis shows the minimal separation $r_{min}/a$ achieved by the inner binary throughout its evolution, and the x-axis is the initial value of $|\jzeffmean|$ (the absolute value is taken for convenience, since $\jzeffmean$ can be also negative). Red color means that $\jzeff$ of that system crossed zero at some point during its evolution; Blue color means $\jzeff$ did not cross zero\footnote{\label{foot:color}During simulations we compute $\jzeff$ only at apocenters. This means that in rare cases some 'crossings' of $\jzeff$ might have been missed (blue points that should be red)}. The three example systems from Figure~\ref{fig:evolution} are shown here as black dots. The black line is the lower limit for the separation of the inner binary from the Lidov-Kozai theory (eq.~\ref{eq:rmin_lk}).

As can be seen in Fig.~\ref{fig:rperi_vs_jzeff}, systems whose $\jzeff$ did not cross zero (blue) roughly follow the predictions of the LK model (black line) with a tail of systems reaching smaller separations (due to the lower values of $|\jzeff|$ obtained throughout the integration). On the other hand, most systems whose $\jzeff$ crossed zero (red) reach much closer separations. In such cases, the value for the minimal separation of the inner binary is a stochastic parameter. Assuming that each system scans its phase space uniformly (\cite{katz_rate_2012}, section II.B), we expect that after $2\cdot10^6$ inner periods, systems would reach separations of $r/a\sim1.5\cdot10^{-7}$. As can be seen in the figure, most systems where $\jzeff$ crossed zero reached such small separations. It is therefore likely that for longer integrations, the minimal separation obtained by these systems will decrease linearly with time. Note that the median is somewhat lower than the expected value for $2\cdot10^6$ orbits, indicating that the distribution of pericenters is not strictly uniform. Further analysis is beyond the scope of this paper.

\section{A criterion for extreme close approaches in hierarchical triple systems with equal mass binaries}
\label{sec:ca_criterion}
In this section we derive an approximate analytic criterion for $\jzeff$ to cross zero (allowing extreme close approaches) for stable triple systems in which the inner binary has equal masses ($m_1=m_2$).

\subsection{Estimating the maximal fluctuations in \texorpdfstring{$\jzeff$}{j z eff}}
\label{sec:jzeff_fluct}

The fluctuations in $\jzeff$ occur within each outer orbit. The angular momentum can be written as a sum of a slowly varying mean $\bar{\textbf{j}}$ which is assumed constant within each outer orbit and the fluctuating part $\Delta \textbf{j}$:

 \begin{equation}
 	\begin{aligned}
      \textbf{j} &= \bar{\textbf{j}} + \Delta \textbf{j} \\
      \textbf{J}_{out} &= \bar{\textbf{J}}_{out} + \Delta \textbf{J}_{out} \\
    \end{aligned}
 \end{equation}

The fluctuations have been calculated to linear order by \cite{luo_double-averaging_2016}. Note that while the long term corrections in \cite{luo_double-averaging_2016} are restricted to the test particle limit, the short term corrections are not.
The mean values of the vectors $\bar{\textbf{j}}$ and $\bar{\textbf{e}}$ can be calculated from the instantaneous values of $\textbf{j}$, $\textbf{e}$ and the outer true anomaly $f_{out}$ using equation (31)  from \cite{luo_double-averaging_2016} (see also appendix B there). 

The fluctuations of the outer orbit's angular momentum can be obtained from those of the inner orbit using the conservation of the total angular momentum, $\Delta \textbf{J}_{out} = - \Delta \textbf{J}$, or in normalized form (see eq.~\ref{eq:j_normalized}):

\begin{equation}
	\Delta \textbf{J}_{out} = - \Jcirc \Delta \textbf{j}
\end{equation}

Expanding $\jzeff$ (eq.~\ref{eq:jzeff}) to first order in $\Delta \textbf{j}$ yields the fluctuating term in $\jzeff$:

 \begin{equation}
   \begin{aligned}
     \jzeff &= \jzeffmean + \Delta \jzeff \ \ \ \ \ \text{where, }  \\
     \jzeffmean &= \bar{\textbf{j}} \cdot \hat{\bar{\textbf{J}}}_{out} + \bar{j}^{2}\frac{\Jcirc}{2\bar{J}_{out}} \\
     \Delta \jzeff &= \left(1+\frac{\jzeffmean \Jcirc}{\bar{J}_{out}}\right) \hat{\bar{\textbf{J}}}_{out} \cdot \Delta \textbf{j} 
   \end{aligned}
   \label{eq:jzeff_djzeff}
 \end{equation}

The mean values hardly change in one outer orbit, so we can set a coordinate system of $\hat{z}$ along the direction of the mean outer angular momentum $\hat{\bar{\textbf{J}}}_{out}$ and $\hat{x}$ along the direction of the outer eccentricity vector $\hat{\bar{\textbf{e}}}_{out}$. In that case eq.~\ref{eq:jzeff_djzeff} becomes:
   
 \begin{equation}
   \Delta \jzeff = \left(1+\frac{\jzeffmean \Jcirc}{\bar{J}_{out}}\right) \Delta j_z
   \label{eq:djzeff_djz}
 \end{equation}

The maximal fluctuation in $j_z$ during a close approach can be obtained using eqs. (35,33,20) from \cite{luo_double-averaging_2016} in the limit $\bar{\textbf{j}} \to 0$ and $\bar{\textbf{e}} \to 1$ yielding:

\begin{equation}
\Delta j_{z,max} = \epsilon_{SA}\left(1 - \bar{e}_{z}^{2}\right) \frac{15}{8}\left(1+\frac{2\sqrt{2}}{3}e_{out}\right)\ , \\
\label{eq:djzmax}
\end{equation}
where $\epsilon_{SA}$ is a dimensionless parameter that sets the scale of the fluctuations and is given by
\begin{equation}
    \esa = \left(\frac{a}{a_{out}}\right)^{\sfrac{3}{2}} \frac{1}{\left(1-e_{out}^{2}\right)^{\sfrac{3}{2}}} \frac{m_{3}}{\sqrt{\left(m_{1}+m_{2}\right)\left(m_{1}+m_{2}+m_{3}\right)}}\ , \\
    \label{eq:esa}
\end{equation}
and $\bar{e}_z^2$ is the projection of $\bar{\textbf{e}}$ onto the direction of $\hat{\bar{\textbf{J}}}_{out}$. The value of $\bar{e}_z^2$ can be obtained from the initial conditions using the following conserved quantity of the double averaged equations which is approximately constant (a consequence of the conservation of the double averaged potential, \cite{lidov_non-restricted_1976}):

\begin{equation}
	C = -\bar{e}^{2}+\frac{5}{2}\bar{e}_z^{2}-\frac{1}{2}\bar{j}_z^{2}\ .
    \label{eq:C}
\end{equation}

The conserved quantity $C$ is computed from the initial conditions using Eq. \eqref{eq:C}.

In the vicinity of close approaches, we can assume the limits $\vec{j}\to0$ and $\vec{e}\to1$ to obtain 
\begin{equation}
		C = -1 + \frac{5}{2} \bar{e}_z^2\ ,
        \label{eq:C_limit}
\end{equation}
which allow $\bar{e}_z$ to be inferred from $C$. 

By calculating $C$ from the initial conditions using Eq. ~\ref{eq:C} and using equations~\ref{eq:C_limit}, \ref{eq:djzmax} ,~\ref{eq:djzeff_djz} we obtain an expression for the maximal value of the fluctuations in $\jzeff$ as a function of the initial conditions:
 
 \begin{equation}
 	\begin{aligned}
 		& \djzeff =& \\
        	& \frac{3}{8}\epsilon_{SA} \left(1+\frac{\jzeffmean \Jcirc}{\bar{J}_{out}}\right) \left(3+2\bar{e}^{2}-5\bar{e}_{z}^{2}+\bar{j}_{z}^{2}\right)\left(1+\frac{2\sqrt{2}}{3}\bar{e}_{out}\right)\Bigg|_{t=0} &
 	\end{aligned}
   \label{eq:djzeff_max}
 \end{equation}
All parameters in eq.~\ref{eq:djzeff_max} are obtained from the initial conditions of the system. A code for calculating the maximal fluctuations using this equation is provided and described in section ~\ref{sec:sneia_implications}.

\subsection{The criterion}
\label{sec:the_criterion}
\begin{figure}
	\includegraphics[width=\columnwidth]{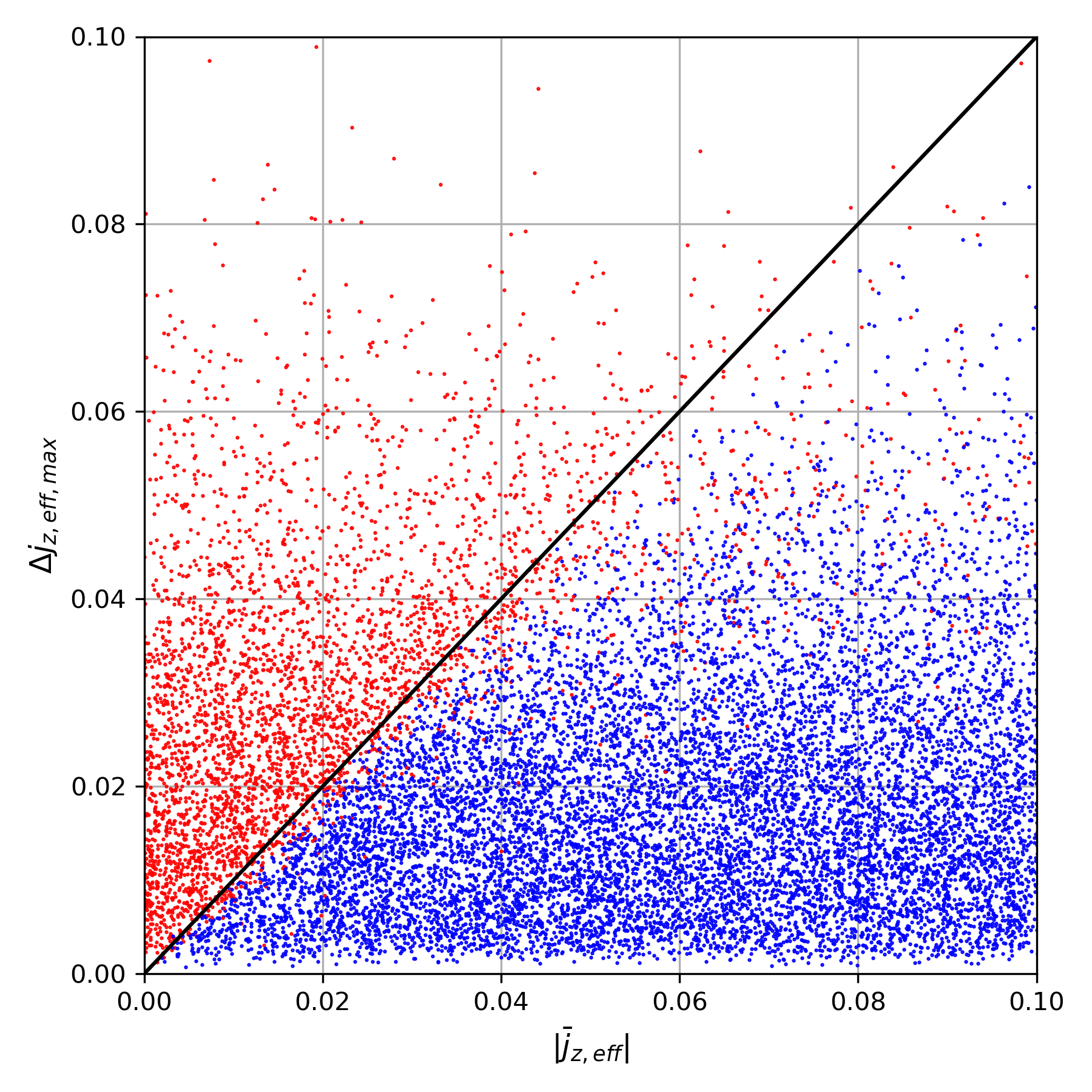}
    \caption{Comparison of individual simulations to the analytic criterion for $\jzeff$ to cross zero, for triple systems with equal mass inner binaries (Eq. \ref{eq:djzeff_gt_jzeffmean}). The dots and colors represent the same systems as in the top panel of Fig. \ref{fig:rperi_vs_jzeff} and are similarly restricted to hierarchies $5\leq\hierarchy\leq10$. For each system, the predicted maximal fluctuation in $\jzeff$ ($\djzeff$), is plotted against the initial mean value of $\jzeff$ ($\jzeffmean$), as calculated from the initial conditions using Eqs.~\ref{eq:jzeff_djzeff} and ~\ref{eq:djzeff_max}. Systems above (below) the black line $\djzeff=\jzeffmean$, are analytically predicted to have $\jzeff$ cross (do not cross) zero. To a good approximation the analytic criterion $\djzeff>\jzeffmean$ (Eq. \ref{eq:djzeff_gt_jzeffmean}) predicts such crossings and therefore extreme close approaches (see Fig~\ref{fig:rperi_vs_jzeff}).}
    \label{fig:djzeff_vs_jzeff}
\end{figure}

As shown in section~\ref{sec:numerical_example}, extreme close approaches occur when $\jzeff$ crosses zero. Using the approximations described in section~\ref{sec:ca_criterion} the condition that $\jzeff$ crosses zero is that 
\begin{equation}
	\djzeff > \jzeffmean\ \ \ (\textrm{at}\ t=0)
    \label{eq:djzeff_gt_jzeffmean}
\end{equation}
where $\jzeffmean$ and $\djzeff$ are calculated from the initial conditions (an implementation in code is provided in section~\ref{sec:sneia_implications}). A comparison of the analytic criterion \eqref{eq:djzeff_gt_jzeffmean} is compared to the results of numerical simulations in Figure~\ref{fig:djzeff_vs_jzeff}. Each dot in the figure represents the results of a simulation of a triple system of masses $m_1=m_2=0.6M_{\sun}$ and $0.1<m_3<1.2M_{\sun}$ and hierarchies $\hierarchy=5,6,7,8,10$. The other initial conditions are sampled from the distributions described in section~\ref{sec:numerics}. Red (blue) dots correspond to systems in which $\jzeff$ crossed (did not cross) 0. As can be seen, the criterion eq.~\eqref{eq:djzeff_gt_jzeffmean} succeeds at predicting whether or not $\jzeff$ would cross zero during its evolution to an excellent approximation. Systems whose $\jzeff$ crossed zero would achieve extreme close approach of the inner binary, as confirmed in Figure~\ref{fig:rperi_vs_jzeff}.

In Figs ~\ref{fig:rates_per_mp} and ~\ref{fig:col_rates_octupole} the analytic fractions, based on the criterion ~\ref{eq:djzeff_gt_jzeffmean}, are shown (labeled analytical). The analytic fractions are obtained by sampling initial conditions for 500,000 triple systems (using the same distribution as the simulations, described in section~\ref{sec:numerics}), and checking for each system whether or not the initial conditions meet the criterion. As seen in the figures, the criterion allows a good estimate to the close approaches fraction for triples with equal mass inner binaries at above-moderate hierarchies $\hierarchy\gtrsim 5$.

For systems with small hierarchies, the system may be quickly disrupted before a close approach in the binary is attained. In these cases the analytic criterion, which is derived based on the assumption that the systems stay intact forever, predicts a close approach fraction which is too high, as can be seen in Fig.~\ref{fig:col_rates_octupole} at low hierarchies. For systems with unequal mass binaries, the octupole term adds another oscillation to $\jzeff$ (see figure \ref{fig:evolution_octupole} and discussion in the text) thus increasing its chance to cross zero. In these cases the fraction predicted based on the analytic criterion is too low. Estimating the fluctuations in $\jzeff$ due to the octupole term is beyond the scope of this paper. However, for this case, criterion~\ref{eq:djzeff_gt_jzeffmean} can be used to compute a lower bound to the close approaches fraction.

\subsection{Implementation of the criterion in computer code}
\label{sec:criterion_mplementation}

Criterion~\ref{eq:djzeff_gt_jzeffmean} can be computed from the initial conditions of a triple system by computing: $\djzeff$ from equation~\ref{eq:djzeff_max}, $\jzeffmean$ from equation~\ref{eq:jzeff_djzeff}, and $\bar{\textbf{j}}$ from equations (20), (31) and appendix B in \cite{luo_double-averaging_2016}. All the parameters and the criterion can be computed using a \textit{Python} file which is attached to this paper.

\section{Summary and Discussion}
\label{sec:sneia_implications}
\subsection{Summary}
It was shown by \cite{katz_rate_2012} that triple systems consisting of an equal-mass inner white-dwarf binary and a comparable mass tertiary at intermediate hierarchies $3\leq\hierarchy\leq10$ have a few percent chance of resulting in extreme close approaches $r/a\lesssim 10^{-5}$ and the collision of the white-dwarfs. In this paper we reproduced and significantly extended the study of the conditions for such extreme close approaches, both analytically and numerically using a new code written by N. Haim which is now publicly available (see \S\ref{sec:numerics}). As described in section~\ref{sec:direct_integration} we extended the numerical experiments for a wide range of inner and tertiary masses. We showed that extreme close approaches are possible for low mass M-dwarf tertiaries (down to  at least $0.1 \rm M_{\odot}$) but with a declining probability that is roughly linear with the mass (see figures \ref{fig:rates_per_mp},\ref{fig:col_rates_octupole}). For equal mass inner binaries in stable systems (usually $\hierarchy\geq5$), we derived in \S\ref{sec:the_criterion} an analytic criterion (Equations \ref{eq:djzeff_max},\ref{eq:djzeff_gt_jzeffmean}, with an implementation provided in \S\ref{sec:criterion_mplementation}) that allows extreme close approaches to be predicted from the initial conditions with good confidence (see Figs \ref{fig:rperi_vs_jzeff},\ref{fig:djzeff_vs_jzeff} for individual runs and Figs \ref{fig:rates_per_mp},\ref{fig:col_rates_octupole} for statistical comparisons). To achieve this, the short-term (outer period) fluctuations of the orbital parameters were calculated using the results of \citep{luo_double-averaging_2016}. Finally, we explored a wide range of unequal mass inner binaries and demonstrated that the collision probability increases linearly as a function of the octupole mass coefficient $(m_2-m_1)/(m_2+m_1)$ (see figure \ref{fig:col_rates_octupole}).

\subsection{Implications for the collision model of type Ia supernovae}

In principle, any close approach of two white dwarfs may lead to interaction. However, in cases of a slowly decreasing inner binary pericenter, energy dissipation or a grazing encounter may affect the orbits and suppress the chances of experiencing a direct collision. It is therefore useful to consider \textit{clean collisions} \citep{katz_rate_2012}, in which all close passages prior to the collision itself are sufficiently far to ignore such effects. In other words, in all early close passages, the separation of the inner binary is always larger than a factor $r>R_{dissip}$. Only in its last passage the separation becomes small enough to allow for a collision $r<R_{col}$. 

\begin{figure}
	\includegraphics[width=\columnwidth]{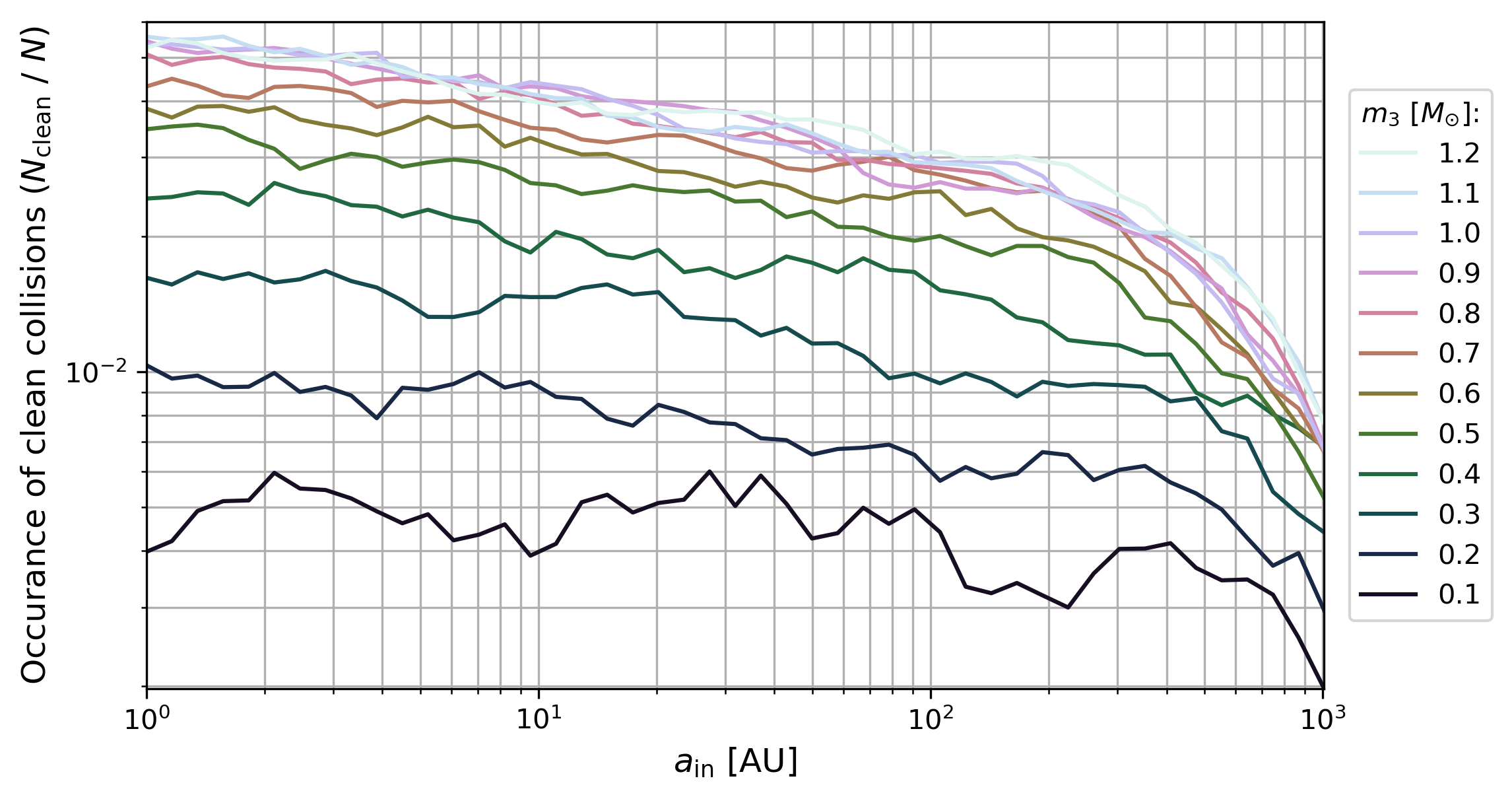}
    \caption{Probability for clean collisions in systems with equal mass inner binaries. These are the same ensembles as the equal mass $m_1=m_2=0.6 M_{\odot}$ case in Fig . \ref{fig:col_rates_octupole}, assuming log-uniform distribution of the hierarchy restricted to the range $3 \leq \hierarchy \leq 10$, and counting only clean collisions where all approaches preceding the collision had separations larger than $R_{dissip}=4R_{WD}$ (see text). 
    \label{fig:clean_collisions_equal}}
\end{figure}

\begin{figure}
	\includegraphics[width=\columnwidth]{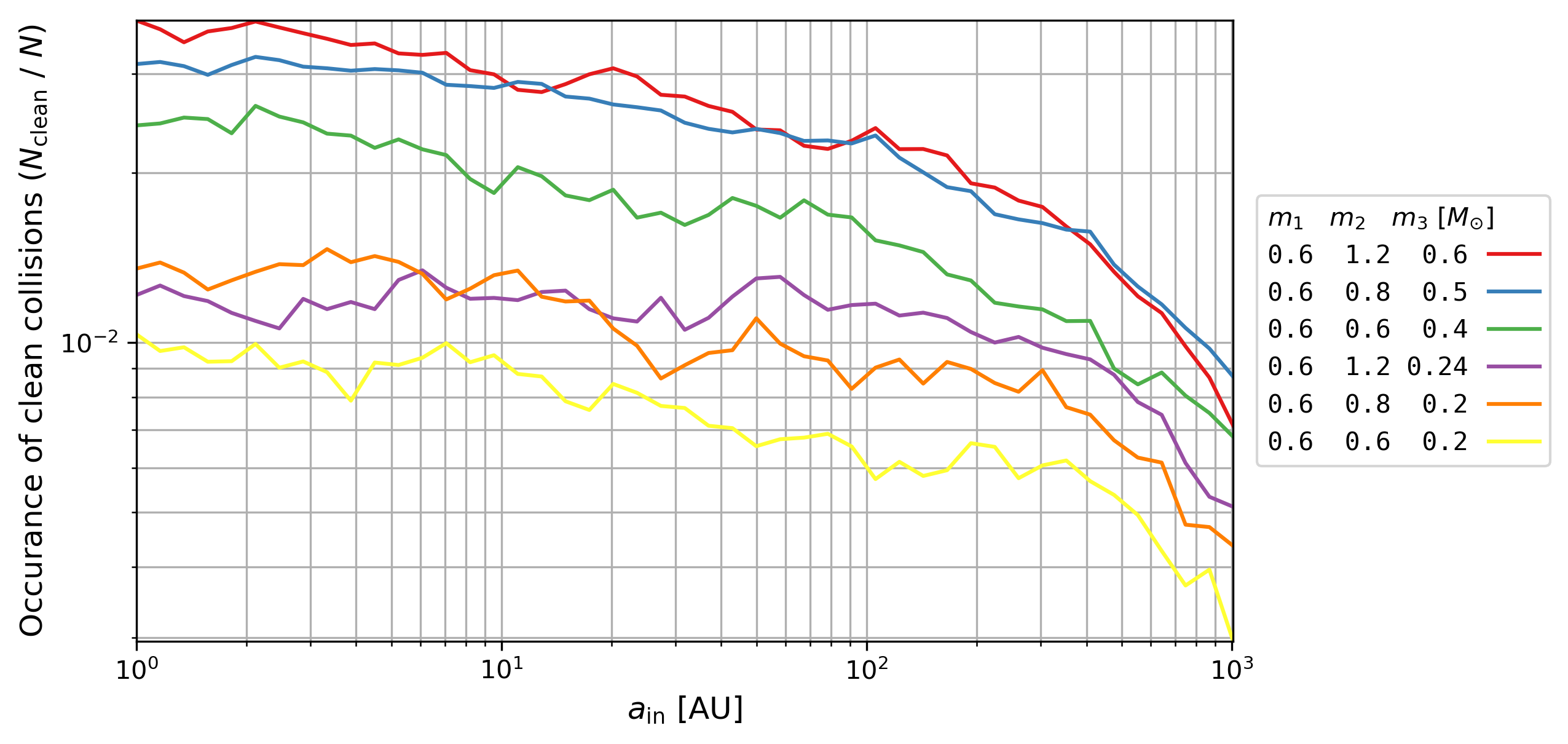}
    \caption{Probability for clean collisions in  systems with unequal mass inner binaries. Same as Fig.\ref{fig:clean_collisions_equal} but for the systems with inner binaries having unequal masses (values of $m_1,m_2,m_3$ are shown in the legend). Two ensembles with equal-mass inner binaries are included to allow for comparison. 
    \label{fig:clean_collisions_unequal}}
\end{figure}

In figures~\ref{fig:clean_collisions_equal} and \ref{fig:clean_collisions_unequal} the clean collisions probabilities are shown for different masses and a wide range of inner-semi major axis $1 < a < 1000$AU, assuming a log-uniform distribution of the initial hierarchy in the range $3\leq\hierarchy\leq10$ \citep[similar to][see section~\ref{sec:numerics} for implementation details]{katz_rate_2012}. In our implementation for clean collisions we use $R_{col}=2R_{WD}$ and $R_{dissip}=4R_{WD}$, where $R_{WD}=10^9cm$. 

These results tighten the constraints on the collision model as a primary channel for type Ia supernovae. About 0.01 of WDs need to explode within a Hubble time to account for the SN Ia rate \citep[e.g. 0.001 type Ia and 0.1 WD per solar mass of star fomration][]{maoz_star_2017}. Given that only $10\%$ of WDs are likely to have a (lighter, wide-orbit) companion WD \citep{klein_way_2017}, collision probabilities in the relevant systems need to be at least $\sim 10\%$. As can be seen in figures \ref{fig:clean_collisions_equal} and \ref{fig:clean_collisions_unequal}, the probability seems to be too low by a factor of few for $\sim M_{\odot}$ tertiaries and an order of magnitude too low for low mass tertiaries $M\lesssim 0.2 M_{\odot}$. The problem may be much worse if only a small fraction of double WDs have tertiaries and if the relevant high inclination systems lead to interaction before the stars become WDs \citep[e.g.][]{toonen_rate_2017}. 

In order to estimate the collision rate, the multiplicity of white-dwarfs needs to be reliably measured. Massive main sequence tertiaries $M\gtrsim 0.5 M_{\sun}$, for which the collision probabilities are higher, are bright $M_V<10, M_K\lesssim 6$ \citep[e.g.][]{Benedict_mass_luminosity_2016} and can be detected with high completeness by adaptive optics surveys of intermediate mass stars \citep[the main progenitors of WDs, e.g.][]{de_rosa_vast_2014} or as common-proper motion companions to A-stars and WDs \citep[e.g.][and soon with Gaia]{de_rosa_vast_2014, farihi_wd_companions_2005}. Direct observational constraints can thus be obtained on such triple systems but this is beyond the scope of this paper. Finaly, we note that while the results of this paper suggest that the collision probabilities in triple systems with low-mass tertiaries are low, the role of low-mass tertiaries in enhancing collisions in higher multiplicity systems  \citep[e.g.][]{pejcha_quadruple_2013,fang_dynamics_2017} remains to be explored.

\section*{Acknowledgements}

We thank Liantong Luo, Doron Kushnir, Subo Dong, Andrew Gould, Silvia Toonen, and Hagai Perets for useful discussions. This research was supported by the ICORE Program (1829/12) and the Beracha Foundation. The computations were performed with a high performance computing facility which is partly supported by the Israel Atomic Energy Commission - The Council for Higher Education - Pazi Foundation and partly by a research grant from The Abramson Family Center for Young Scientists.




\bibliographystyle{mnras}
\bibliography{refs} 



\appendix

\section{Relation of \texorpdfstring{$\jzeff$}{j z eff} to the mutual inclination}
\label{app:jzeff_inclination}
In the test particle approximation, the condition $\jzeff=0$ reduces to $j_z=0$ or a mutual inclination between the inner and outer orbit of $90$ degrees.
It is therefore useful to express the relation between $\jzeff$ and the mutual inclination $i$ in the non-test particle case. Eq.~\ref{eq:jzeff} can be expressed as:

 \begin{equation}
    \jzeff = j \left( \cos{i} + j \frac{\Jcirc}{2\lvert J_\text{out}\rvert} \right)\ .
\end{equation}

The condition $\jzeff=0$, where close approaches are expected, requires initial retrograde configurations $i>90$ as can be inferred also by the equivalent condition $|J_{tot}|=|J_{out}|$ \citep[Eq. \ref{eq:J_bound}][]{lidov_non-restricted_1976}.

\section{Integration method}
\label{app:integration}

 Similar to \citep{katz_rate_2012} we use the PTMT, second-order, symplectic integrator with adaptive time-step \citep{preto_class_1999,mikkola_algorithmic_1999}. The propagation of the positions and velocities of the bodies $\textbf{x}$ and $\textbf{v}$ to the next time step is performed in a two stage leapfrog scheme:
 
 \begin{equation}
 	\begin{aligned}
    	\textbf{v}_{next} &=\textbf{v} + \textbf{a}(\textbf{x}) \cdot dt_0 \cdot \left( \frac{U(\textbf{x})}{U_0} \right)^{-3/2} \\
 		\textbf{x}_{next} &=\textbf{x} + \textbf{v}_{next} \cdot dt_0 \cdot \left( \frac{E_0-K(\textbf{v}_{next})}{U_0} \right)^{-3/2} \\
 	\end{aligned}
 \end{equation}
Where $\textbf{a}$ are the accelerations, $E_0$ is the initial total energy of the system, $K$ and $U$ are the  kinetic and potential energies respectively and $dt_0$ and $U_0$ are constants. $U_0$ is chosen as ${U_0 = -Gm_1m_2 / a_0}$ where $a_0$ is the initial semi major axis. The time-step amplitude is parametrized as $dt_0=P_{in,0}/N_s$, where $P_{in,0}$ is the initial period of the inner binary and for most of our runs we use $N_s=1000$. For the convergence check (section~\ref{app:convergence}) we also use $N_s=250,50$.

\section{Convergence}
\label{app:convergence}

\begin{figure}
	\includegraphics[width=\columnwidth]{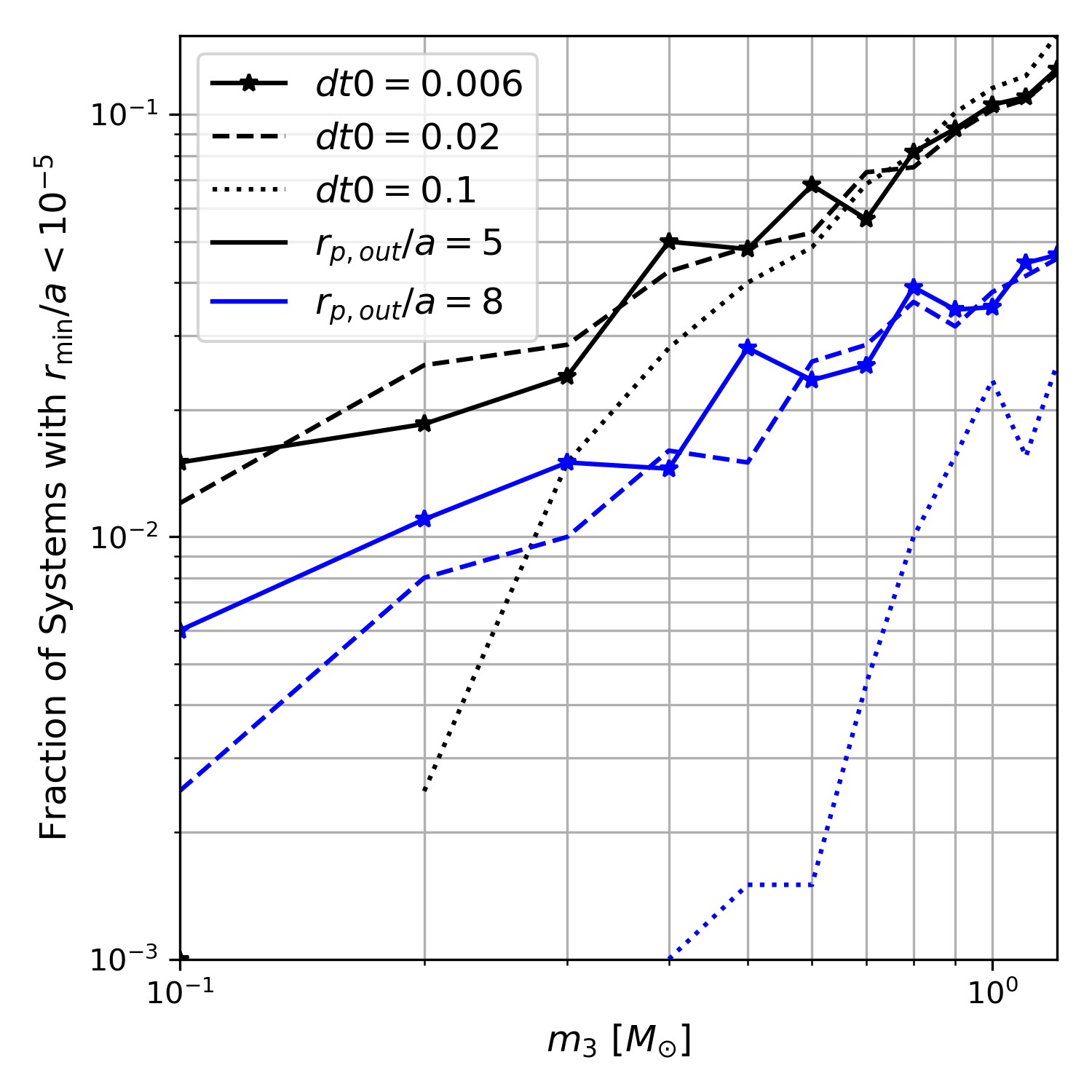}
    \caption{Convergence test. Each dot represents a system of masses $m_1=0.6,m_2=0.6,m_3=0.1-1.2M_{\sun}$ and initial hierarchy $\hierarchy=5,8$ (black, blue resp.). Collision probabilities are shown for $dt0=0.006,0.02,0.1$ (solid, dashed, dotted resp.). As can be seen, the results shown in this paper (for simulations with $dt0=0.006$) are converged to within statistical error.}
    \label{fig:convergence}
\end{figure}

In Figure~\ref{fig:convergence} we show the close approaches probability (the fraction of systems, out of the total systems for the same parameters, whose inner binary separation became smaller than the semi major axis by $10^{-5}$) for triple systems with masses $m_1=m_2=0.6M_{\sun}$ and $0.1 < m_3 < 1.2 M_{\sun}$. We show the results for runs with different time-step amplitudes with $N_s=1000,250,50$ shown in solid, dashed and dotted lines respectively, see appendix~\ref{app:integration}), and for two initial hierarchies $r_{p,out}=5,8$. The solid lines are the same lines as in Figure~\ref{fig:rates_per_mp} for the relevant parameters. As can be seen fast convergence is achieved as a function of $N_s$.

\begin{figure}
	\includegraphics[width=\columnwidth]{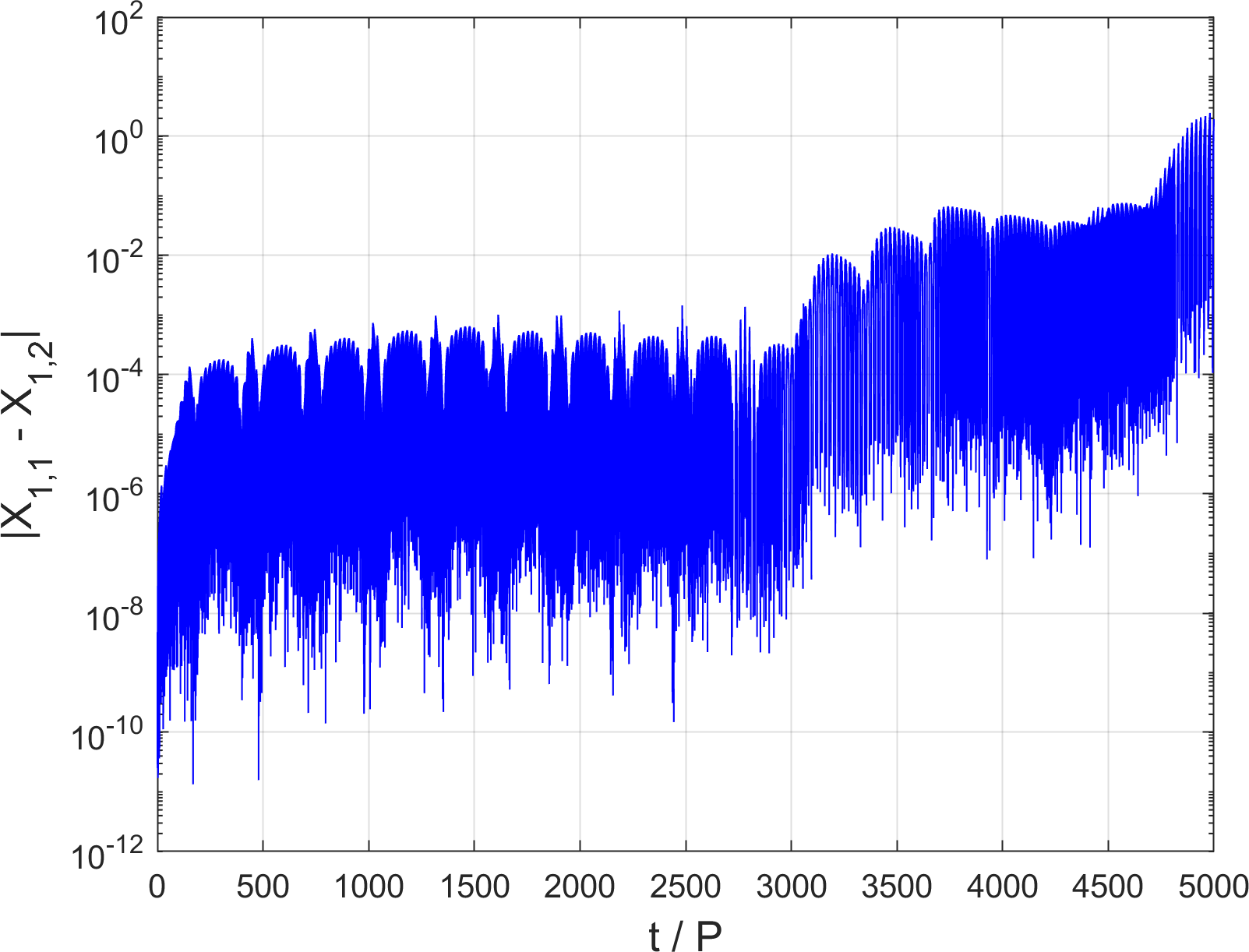}
    \caption{A short Lyapunov time demonstration. Evolution of the difference in the x-coordinate of $m_1$ in two integrations with the same initial conditions except for an initial difference of $10^{-8}$ in the value for this coordinate. As can be seen, after about $\sim$5,000 inner periods the difference between the two systems grew to order unity. This implies an average Lyapunov timescale of a few hundreds of inner orbits.}
    \label{fig:lyapunov}
\end{figure}

It is important to note that most of the simulations are not individually converged. In figure~\ref{fig:lyapunov} the results of two integrations are shown which are identical in all parameters except for position of $m_1$ along the x-axis, which was modified by $10^{-8}a$. The figure shows the evolution of the difference between the positions of $m_1$ along the x-axis in the two integrations. As can be seen, after a few thousand inner periods, the difference grows to order unity. The Lyapunov timescale for this system is thus of order a few hundreds of inner orbits on average. In our simulations we integrate each system to millions of inner periods, and it is therefore futile to aim for the convergence of an individual integration. Having said that, As can be seen in Fig.~\ref{app:convergence}, The statistical properties of the systems, such as the close approaches probabilities, are indeed converged to within statistical error. 

Another issue with the integrator is that for some systems with very close approaches the time-step became so small that the integration was not completed. For the small-time step amplitudes that we used with $N_s=1000$, this happened for only a handful of systems (out of $\sim$100,000) and we did not consider them in the results. For cruder time resolutions (smaller $N_s$), this happens more frequently. For our convergence tests, out of a total of $\sim$10,000 runs for each $N_s$, the number of such cases were 70 for $N_s=250$ and 910 for $N_s=50$. However, even for $N_s=50$, where 10\% of the runs were stuck, the overall convergence statistics that we show is not significantly affected.

\section{Summary of the properties of the numerical Ensembles}
\label{app:ensembles}
\begin{figure}
	\includegraphics[width=\columnwidth]{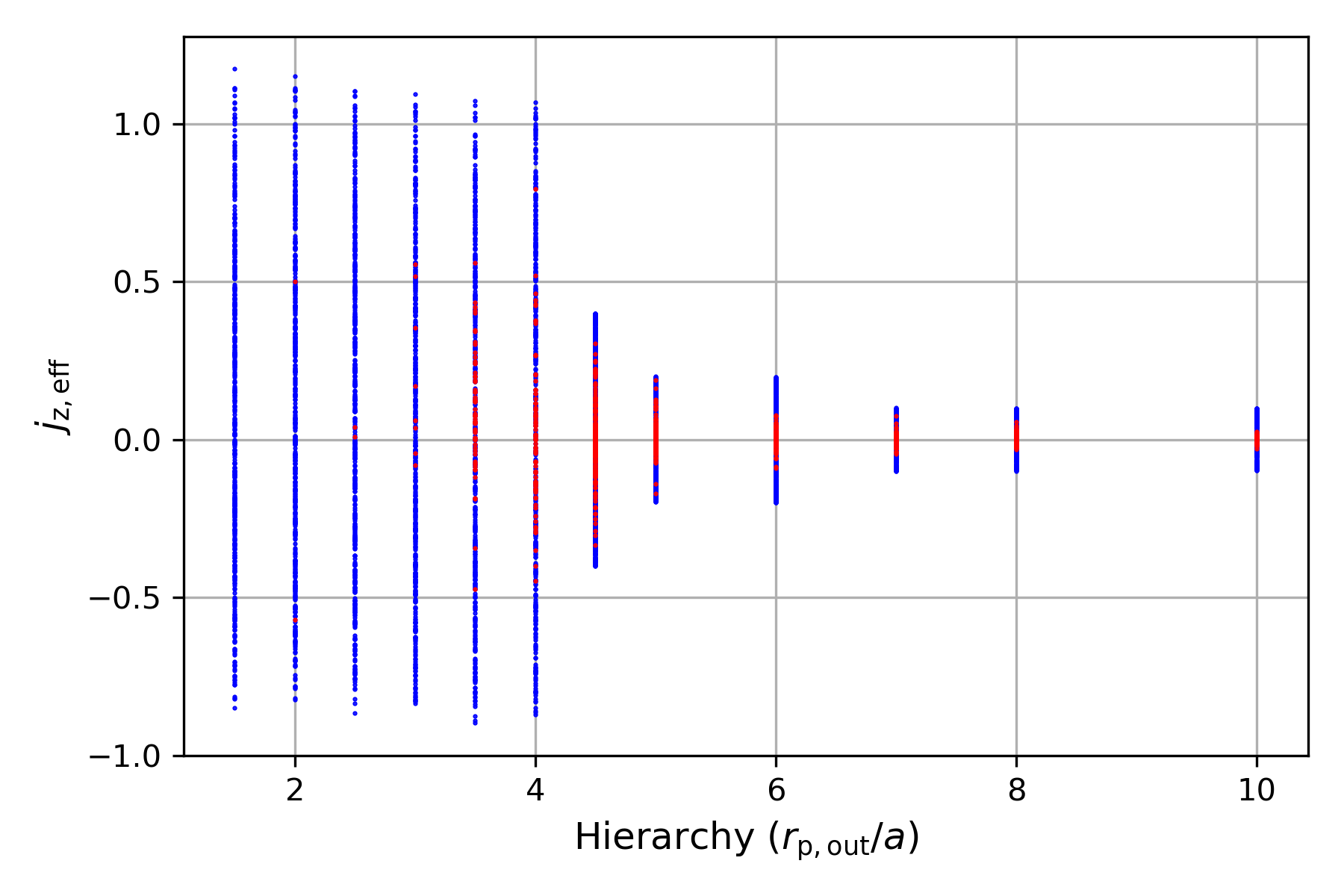}
    \caption{Examples of ensembles with and without cuts in $\jzeff$. Each dot represents a three body system of masses $m_1=0.6,m_2=0.6,m_3=0.8M_{\sun}$. For each system we show the initial $\jzeff$ plotted against the initial hierarchy $\hierarchy$. Systems whose inner binary experienced (did not experience) a close approach of $r/a < 10^{-5}$ are shown in red (blue). For small hierarchies ($\hierarchy \leq 4$) all sampled initial conditions are simulated whereas for $\hierarchy \geq 4.5$ simulations are performed only for initial conditions with $|\jzeff|$ smaller than a predefined maximum $|\jzeff|_{\max}$ as can be seen by the limited range of $\jzeff$ in these hierarchies. As can be seen, the vast majority of systems that experienced a close approach are concentrated near initial $\jzeff=0$ and are far from the edges of the simulated range, justifying the cuts.}
    \label{fig:jzeff_cut}
\end{figure}

A summary of all numerical ensembles that were calculated in this work is given in table \ref{tab:sim_ensembles}.
For low hierarchies $1.5\leq\hierarchy\leq4$ all $N_\text{batch}$=500 sampled initial conditions are integrated. For higher hierarchies $4.5\leq\hierarchy\leq10$, $N_\text{batch}$=2000 initial conditions are sampled but systems with large initial $|\jzeff|$ are assumed to not lead to close approaches and only a subsample satisfying a prescribed upper limit $|\jzeff|<|\jzeff|_{\max}$ are simulated (the values of $|\jzeff|_{\max}$ are given in table \ref{tab:batches_cut_at} and the number of simulations actually performed is given in table \ref{tab:sim_ensembles}). The range of $|\jzeff|$ is empirically chosen so that the vast majority of the systems that experience close approaches lie deep within the region of $|\jzeff| < |\jzeff|_{\max}$. In Figure~\ref{fig:jzeff_cut} the initial values of $|\jzeff|$ for the systems for different initial hierarchies $\hierarchy$ are shown. Systems that  experienced (did not experience) a close approach of $r/a<10^{-5}$ are shown in red (blue). As can be seen, most systems that experience a close approach have small initial $\jzeff=0$. Using such plots we verify that the chosen values $|\jzeff|_{\max}$ for each ensemble captures the majority of systems that experience a close approach. We emphasize that this is a conservative approach since all systems that are not simulated are assumed to not lead to a close approach.

\begin{table*}
  \caption{Numerical ensembles. For each choice of mass values (first three columns) and initial hierarchy $\hierarchy$ (first row), we show the number of systems that were eventually integrated. The total number of systems integrated is 120,046.}
  \label{tab:sim_ensembles}
  \begin{tabular}{lllllllllllllll}
  \toprule
    m1 &  m2&m3 $~~~\hierarchy=$ &  1.5 &  2.0 &  2.5 &  3.0 &  3.5 &  4.0 &   4.5 &   5.0 &  6.0 &  7.0 &  8.0 & 10.0 \\
  \midrule
   0.6 &  0.6 &              0.1 &  500 &  500 &  500 &  500 &  500 &  500 &   531 &   489 &  533 &  238 &  251 &  267 \\
   0.6 &  0.6 &              0.2 &  500 &  500 &  500 &  500 &  500 &  500 &   500 &   479 &  516 &  244 &  270 &  245 \\
   0.6 &  0.6 &              0.3 &  500 &  500 &  500 &  500 &  500 &  500 &   464 &   476 &  495 &  267 &  244 &  256 \\
   0.6 &  0.6 &              0.4 &  500 &  500 &  500 &  500 &  500 &  500 &   497 &   506 &  464 &  251 &  250 &  258 \\
   0.6 &  0.6 &              0.5 &  500 &  500 &  500 &  500 &  500 &  500 &   507 &   498 &  475 &  248 &  276 &  249 \\
   0.6 &  0.6 &              0.6 &  500 &  500 &  500 &  500 &  500 &  500 &   757 &   514 &  501 &  256 &  248 &  252 \\
   0.6 &  0.6 &              0.7 &  500 &  500 &  500 &  500 &  500 &  500 &   891 &   479 &  486 &  260 &  234 &  260 \\
   0.6 &  0.6 &              0.8 &  500 &  500 &  500 &  500 &  500 &  500 &   990 &   503 &  525 &  247 &  254 &  224 \\
   0.6 &  0.6 &              0.9 &  500 &  500 &  500 &  500 &  500 &  500 &  1361 &   775 &  497 &  246 &  260 &  260 \\
   0.6 &  0.6 &                1 &  500 &  500 &  500 &  500 &  500 &  500 &  1364 &   774 &  517 &  252 &  230 &  243 \\
   0.6 &  0.6 &              1.1 &  500 &  500 &  500 &  500 &  500 &  500 &  1348 &   982 &  486 &  264 &  248 &  251 \\
   0.6 &  0.6 &              1.2 &  500 &  500 &  500 &  500 &  500 &  500 &  1322 &   951 &  513 &  269 &  251 &  215 \\
   0.6 &  1.2 &              0.6 &  500 &  500 &  500 &  500 &  500 &  500 &  1228 &  1261 &  735 &  497 &  512 &  522 \\
   0.6 &  1.2 &              0.3 &  500 &  500 &  500 &  500 &  500 &  500 &   738 &   487 &  490 &  236 &  226 &  265 \\
   0.6 &  0.9 &              0.5 &  500 &  500 &  500 &  500 &  500 &  500 &   860 &   738 &  536 &  267 &  245 &  235 \\
   0.6 &  0.9 &             0.25 &  500 &  500 &  500 &  500 &  500 &  500 &   752 &   511 &  497 &  247 &  242 &  231 \\
   0.6 &  0.7 &  $\frac{1.3}{3}$ &  500 &  500 &  500 &  500 &  500 &  500 &   910 &   726 &  482 &  240 &  232 &  248 \\
   0.6 &  0.7 &  $\frac{1.3}{6}$ &  500 &  500 &  500 &  500 &  500 &  500 &   861 &   771 &  501 &  253 &  251 &  243 \\
   0.6 &  1.8 &              0.8 &  500 &  500 &  500 &  500 &  500 &  500 &  1197 &  1249 &  729 &  497 &  479 &  494 \\
   0.6 &  1.8 &              0.4 &  500 &  500 &  500 &  500 &  500 &  500 &   912 &   715 &  636 &  630 &  604 &  624 \\
  \bottomrule
  \end{tabular}
\end{table*}

\begin{table*}
  \caption{Maximal initial values of $|\jzeff|$ for which simulations were performed. For each mass choice (first three columns) and initial hierarchy $\hierarchy$ (first row), the predefined maximal value $|\jzeff|_{\max}$ is given. Only initial conditions satisfying $|\jzeff|<|\jzeff|_{\max}$ are integrated, while the rest are assumed to not lead to a close approach.}
  \label{tab:batches_cut_at}
  \begin{tabular}{lllllllllllllll}
  \toprule
    m1 &   m2 m3 $~~~\hierarchy=$& 1.5 & 2.0 & 2.5 & 3.0 & 3.5 & 4.0 &   4.5 &  5.0 &   6.0 &   7.0 &   8.0 &  10.0 \\
  \midrule
   0.6 &  0.6 &              0.1 &   - &   - &   - &   - &   - &   - &   0.2 &  0.2 &   0.2 &   0.1 &   0.1 &   0.1 \\
   0.6 &  0.6 &              0.2 &   - &   - &   - &   - &   - &   - &   0.2 &  0.2 &   0.2 &   0.1 &   0.1 &   0.1 \\
   0.6 &  0.6 &              0.3 &   - &   - &   - &   - &   - &   - &   0.2 &  0.2 &   0.2 &   0.1 &   0.1 &   0.1 \\
   0.6 &  0.6 &              0.4 &   - &   - &   - &   - &   - &   - &   0.2 &  0.2 &   0.2 &   0.1 &   0.1 &   0.1 \\
   0.6 &  0.6 &              0.5 &   - &   - &   - &   - &   - &   - &   0.2 &  0.2 &   0.2 &   0.1 &   0.1 &   0.1 \\
   0.6 &  0.6 &              0.6 &   - &   - &   - &   - &   - &   - &   0.3 &  0.2 &   0.2 &   0.1 &   0.1 &   0.1 \\
   0.6 &  0.6 &              0.7 &   - &   - &   - &   - &   - &   - &  0.35 &  0.2 &   0.2 &   0.1 &   0.1 &   0.1 \\
   0.6 &  0.6 &              0.8 &   - &   - &   - &   - &   - &   - &   0.4 &  0.2 &   0.2 &   0.1 &   0.1 &   0.1 \\
   0.6 &  0.6 &              0.9 &   - &   - &   - &   - &   - &   - &   0.4 &  0.3 &   0.2 &   0.1 &   0.1 &   0.1 \\
   0.6 &  0.6 &                1 &   - &   - &   - &   - &   - &   - &   0.4 &  0.3 &   0.2 &   0.1 &   0.1 &   0.1 \\
   0.6 &  0.6 &              1.1 &   - &   - &   - &   - &   - &   - &   0.4 &  0.4 &   0.2 &   0.1 &   0.1 &   0.1 \\
   0.6 &  0.6 &              1.2 &   - &   - &   - &   - &   - &   - &   0.4 &  0.4 &   0.2 &   0.1 &   0.1 &   0.1 \\
   0.6 &  1.2 &              0.6 &   - &   - &   - &   - &   - &   - &   0.5 &  0.5 &   0.3 &   0.2 &   0.2 &   0.2 \\
   0.6 &  1.2 &              0.3 &   - &   - &   - &   - &   - &   - &   0.3 &  0.2 &   0.2 &   0.1 &   0.1 &   0.1 \\
   0.6 &  0.9 &              0.5 &   - &   - &   - &   - &   - &   - &  0.35 &  0.3 &   0.2 &   0.1 &   0.1 &   0.1 \\
   0.6 &  0.9 &             0.25 &   - &   - &   - &   - &   - &   - &   0.3 &  0.2 &   0.2 &   0.1 &   0.1 &   0.1 \\
   0.6 &  0.7 &  $\frac{1.3}{3}$ &   - &   - &   - &   - &   - &   - &  0.35 &  0.3 &   0.2 &   0.1 &   0.1 &   0.1 \\
   0.6 &  0.7 &  $\frac{1.3}{6}$ &   - &   - &   - &   - &   - &   - &  0.35 &  0.3 &   0.2 &   0.1 &   0.1 &   0.1 \\
   0.6 &  1.8 &              0.8 &   - &   - &   - &   - &   - &   - &  0.55 &  0.7 &  0.55 &   0.5 &   0.4 &   0.3 \\
   0.6 &  1.8 &              0.4 &   - &   - &   - &   - &   - &   - &  0.35 &  0.3 &  0.25 &  0.25 &  0.25 &  0.25 \\
  \bottomrule
  \end{tabular}
\end{table*}


\bsp	
\label{lastpage}
\end{document}